# The Nature of Ferromagnetism in the Chiral Helimagnet Cr$_{1/3}$NbS$_2$

N. Sirica[1,2*], P. Vilmercati[1], F. Bondino[3], I. Pis[3,4], S. Nappini[3], S. –K. Mo[5], A. V. Federov[5], P. K. Das[3,6], I. Vobornik[3], J. Fujii[3], L. Li[7], D. Sapkota[1], D. S. Parker[8], D. G. Mandrus[7,8,1] and N. Mannella[1§]

[1]*Department of Physics and Astronomy, The University of Tennessee, Knoxville, Tennessee 37996, USA*
[2]*Center for Integrated Nanotechnologies, Los Alamos National Laboratory, Los Alamos, NM 87545, USA*
[3]*IOM CNR Laboratorio TASC, S.S. 14 Km 163.5, I-34149 Basovizza (TS), Italy*
[4]*Elettra-Sincrotrone Trieste S.C.p.A., S.S. 14 Km 163.5, I-34149 Basovizza (TS), Italy*
[5]*Advanced Light Source, Lawrence Berkeley National Laboratory, Berkeley, CA 94720, USA*
[6]*International Centre for Theoretical Physics, Strada Costiera 11, 34100 Trieste, Italy.*
[7]*Department of Materials Science and Engineering, The University of Tennessee, Knoxville, Tennessee 37996, USA*
[8]*Materials Science and Technology Division, Oak Ridge National Laboratory, Oak Ridge, Tennessee 37831, USA*

[*]nsirica@lanl.gov
[§]nmannell@utk.edu

**ABSTRACT**
The chiral helimagnet, Cr$_{1/3}$NbS$_2$, hosts exotic spin textures, whose influence on the magneto-transport properties, make this material an ideal candidate for future spintronic applications. To date, the interplay between macroscopic magnetic and transport degrees of freedom is believed to result from a reduction in carrier scattering following spin order. Here, we present electronic structure measurements through the helimagnetic transition temperature, T$_C$ that challenges this view by showing a Fermi surface comprised of strongly hybridized Nb- and Cr-derived electronic states, and spectral weight in proximity to the Fermi level to anomalously increases as temperature is lowered below T$_C$. These findings are rationalized on the basis of first principle, density functional theory calculations, which reveal a large nearest-neighbor exchange energy, suggesting the interaction between local spin moments and hybridized Nb- and Cr-derived itinerant states to go beyond the perturbative interaction of Ruderman-Kittel-Kasuya-Yosida, suggesting instead a mechanism rooted in a Hund's exchange interaction.

**Editor Summary**
In the chiral helimagnet Cr$_{1/3}$NbS$_2$, localized spin moments on the Cr site are believed to play a passive role in the material's electronic and transport properties. This interpretation is challenged by the experimental observation of both hybridization between local magnetic moments and itinerant electrons, and changes occurring in the electronic structure with the onset of magnetism.

**Introduction**

The development of next generation electronic devices relies on the ability to precisely control other intrinsic degrees of freedom (DOF) beyond that of charge. In this endeavor, chiral helimagnets (CHMs) have emerged as promising materials to be used for spintronic and other information technology applications, where the intent is to actively manipulate either the individual, or average spin angular momentum of itinerant carriers[1]. In CHMs the chiral framework of the crystal structure allows for the arrangement of spins into incommensurate periodic spirals, or helices, resulting from an anti-symmetric spin interaction known as the Dzyaloshinskii-Moriya interaction[2,3,4,5,6]. Such an interaction, which originates from relativistic spin-orbit coupling, not only underlies the helical ground state of CHMs, but unusual spin textures like the two-dimensional spin vortex known as a skyrmion as well[7,8]. Skyrmions are of particular technological interest thanks to the fact that they can be manipulated at the nanoscale level by externally applied magnetic fields[8], spin polarized currents[9,10], and are found to have a profound influence on electronic transport[11,12,13].

Given these important functionalities, there has been a great incentive for the discovery of new CHM materials relevant to the design and fabrication of future spintronic devices[14]. One such material is $Cr_{1/3}NbS_2$[15], which has been shown to host a unique one-dimensional solitonic excitation known as a chiral soliton lattice (CSL)[16]. Much like skyrmions the CSL can be manipulated by external magnetic field[16], but unlike other CHMs, $Cr_{1/3}NbS_2$ crystallizes in a more anisotropic, $Nb_3CoS_6$ (hp20), layered structure consisting of ferromagnetic (FM) planes of intercalated Cr atoms ordering in a ($\sqrt{3}\times\sqrt{3}$)R(30°) superstructure within the *ab*-plane. Each Cr atom occupies a trigonally distorted octahedral ($O_h$) sites in the van der Waal gaps of $2H$-$NbS_2$ (Fig. 1(a))[17] and hosts a local spin moment of ~ 3 $\mu_B$/Cr that order below a Curie temperature, $T_C$ = 116-132 K[15,16,17]. Due to such high anisotropy and the fact that inversion symmetry breaks only along the non-centrosymmetric *c*-axis, where the long range (~ 48 nm) helimagnetic ordering develops, $Cr_{1/3}NbS_2$ has been recognized as an ideal system for studying spin-textures in magnetic thin films, devices fabricated on a substrate, and spin-orbit coupling effects in magnetic multilayers[18,19].

More generally, $Cr_{1/3}NbS_2$ belongs to a class of transition metal dichalcogenides (TMDs) intercalated by magnetic, 3d elements. According to a conventional description of the electronic structure for these materials, Curie-Weiss spin moments localized on the Cr site do not alter the Fermi surface (FS) or band topology of the host $NbS_2$ compound, aside from raising the chemical potential following a donation of charge to Nb-derived bands[20,21]. Due to the large separation distance between Cr neighbors ($\sqrt{3}a$), long range magnetic order is believed to be mediated by Nb conduction electrons through a Ruderman-Kittel-Kasuya-Yosida (RKKY) exchange interaction[22,23,24,25]. Under such an exchange mechanism, the raise in chemical potential that accompanies intercalation serves to alter the nesting condition in the FS, and in turn change the periodicity and range of interaction for the oscillatory exchange coupling between screened, local spin moments (LSM)[26]. In this regard, the identity of the intercalant atom has little impact on the resultant order or ordering temperature so long as each species possesses the same oxidation state and spacing in the lattice. Hence, both the drop in resistivity and negative magnetoresistance (MR) observed throughout this class of materials are argued to result from a reduction in spin scattering due entirely to magnetic ordering[27,28]. However, recent experimental and theoretical findings for $Cr_{1/3}NbS_2$, not only propose an alternative exchange mechanism beyond a conventional RKKY[29], but show abrupt, anisotropic changes in the magneto-transport that suggest Cr-derived states play a more active role in the interplay between LSMs and itinerant electrons[30].

The importance of these findings is underscored by a major technological appeal of $Cr_{1/3}NbS_2$, which lies in the interplay between transport and magnetic DOF. Namely, inter-layer magneto-transport measurements have revealed a direct correlation between the transverse, negative MR and the soliton lattice density, which is attributed to Bragg scattering of conduction electrons by the magnetic superlattice potential of the CSL[27,28]. However, recent in-plane transport, MR and magnetic anisotropy measurements both in proximity to $T_C$, and at the CHM to CSL transition show abrupt changes[29,31] that depend sensitively on the direction of applied magnetic field relative to the *ab*-plane[30]. This includes a drop in the in-plane resistivity by more than an order of magnitude as $T < T_C$ as well as a decrease in the intra-layer MR by 6% at the CSL transition[29]. Such results cannot be readily explained by spin ordering arguments alone, as it is not obvious how a modulation of the CSL state along the crystallographic *c*-axis could have such a strong impact on the in-plane transport properties. Rather, N. J. Ghimire, et al[29] and A. C. Bornstein, et al[30] argue on the basis of first principle calculations that details of the electronic structure must play a more central role in shaping magneto-transport in $Cr_{1/3}NbS_2$. Thus, there exists an experimental need to better elucidate the microscopic mechanisms underlying such a close interplay between in-plane transport and magnetic DOF in this material.

In this article, we provide experimental evidence showing the electronic structure to play a non-trivial role in the description of in-plane magneto-transport in $Cr_{1/3}NbS_2$. Through use of resonant photoemission spectroscopy (ResPES) we find the FS of this material to be comprised of strongly hybridized Cr- and Nb-derived electronic states, while angle-resolved photoemission spectroscopy (ARPES) reveals an anomalous increase of spectral weight in proximity to the Fermi level ($E_F$) as temperature is lowered below $T_C$. Such behavior is inconsistent with that of a conventional itinerant ferromagnet, and can be rationalized on the basis of first principle density functional theory (DFT) calculations to result from a strong exchange (Hund's) interaction between itinerant electrons and LSMs on the Cr sites.

**Results:**

*Overview of Electronic Structure*—Previous work focusing on how the electronic structure of $NbS_2$ is modified with Cr intercalation for temperatures below the helimagnetic transition temperature (Fig. 1(b)) is reported by N. Sirica, et al[32]. To summarize these results, Fig. 1(c-d) shows a schematic diagram of both the FS and band dispersion along the high-symmetry ΓK axis, where crossing points α, $β_{1,2}$, γ, and δ follow the same labeling convention as used in this previous study[32]. As expected, Cr intercalation injects electronic charge into the $NbS_2$ layers, resulting in a sinking of the bonding, bilayer split $NbS_2$ band (S) below $E_F$, while the remaining $NbS_2$-derived, anti-bonding band has crossings points α and γ consistent with a raise in chemical potential. However, the electronic charge introduced by the Cr intercalant not only changes the total number of carriers, but causes a significant modification to the electronic structure of $NbS_2$ through the addition of two additional bands at Γ ($β_1$, $β_2$), and one additional band at K (δ). Since these bands cannot be rationalized on the basis of a rigid band picture, it is natural to inquire about their origin. Here, we focus on the temperature dependence of the α and $β_{1,2}$ bands in proximity to Γ, as the bands at K are found to exhibit no significant change with temperature across $T_C$.

*Temperature-Dependent Angle-Resolved Photoemission*—Temperature dependent ARPES is used to map the dispersion of the α and $β_{1,2}$ bands in proximity to $E_F$ as the temperature is tuned across $T_C$ (Fig. 2). For T < 131K, three dispersive bands can be identified at Γ, while only two bands are present as temperature is raised above $T_C$. From Fig. 2, it is evident that the outermost $β_2$ band shifts towards Γ with increasing temperature until the two β-bands can no longer be distinguished. Such behavior is more clearly demonstrated in Fig. 3, which gives the temperature dependence of the band crossing points ($k_F$) along the two high symmetry directions ΓM (Fig. 3(a)) and ΓK (Fig 3(b)). Here, momentum distribution curves (MDCs) illustrating the distributions of spectral weight at a fixed energy E = $E_F$ along these two high symmetry directions reveals a monotonic shift in $k_F$ for the $β_2$ band to be accompanied by a progressive decrease in spectral weight for 50K < T < 120K. The fact that this behavior is observed along both the ΓM and ΓK directions suggests it to be isotropic about Γ. Moreover, given the close proximity to $T_C$, the relative shift of $β_2$ towards $β_1$ implies the $β_{1,2}$ bands arise from magnetic exchange splitting. While dedicated spin-resolved ARPES measurements are required to establish unambiguously the spin polarization of these bands, such a finding is consistent with the notion that the $β_{1,2}$ bands result from intercalation of a magnetic element (Cr) into $NbS_2$, and suggest the presence of Cr-derived states at $E_F$. Indeed, this finding has been confirmed with ResPES, a methodology that allows photoemission spectra of the valence band (VB) to acquire elemental sensitivity [33, 34].

*Resonant Photoemission*—By tuning the incident photon energy across the Cr L-edge it is possible to identify Cr 3d states in the VB within 3 eV of $E_F$ (Fig. 4(a-b)). The increase in signal intensity at resonance, i.e. with the photon energy tuned on the maximum of the Cr absorption edge, reveals that the structures at ≈ 2.5 eV and within ≈ 1 eV from $E_F$ are states possessing Cr 3d character.

The states at Γ have been shown by polarization dependent ARPES to exhibit predominantly out-of-plane orbital character[32]. Therefore, in contrast to our previous ResPES measurements[32], whose purpose was to identify non-dispersive Cr states remaining on the surface following sample cleave, the photon polarization used in this report was chosen to emphasize only those bulk states having an out-of-plane character through ensuring a component of the photon polarization to lie perpendicular to the sample plane. In doing so, a direct comparison of ResPES spectra taken with photon energies tuned across the Cr L- (Fig. 4 (a-b)) and Nb M- (Fig. 4(c-d)) edges reveals a clear resonance over the Nb "$d_z^2$ sub-band"[21] ($E_B$ ≈ 0.6 eV < E < $E_F$), indicating the presence of hybridization between Cr- and Nb-derived states in the VB. This is made possible by noting spectra taken across the Cr L-edge to be sensitive, in these experimental conditions, to the projection of Cr 3d orbitals, whose splitting is dictated by a trigonally distorted octahedral environment[22,23]. Hence, our ResPES data clarify that the VB states at Γ originate from a linear combination of Nb $d_z^2$ and mixed Cr 3d orbitals ($\sqrt{2}\,d_{xz} - d_{x^2-y^2}$ and $\sqrt{2}\,d_{yz} + d_{xy}$), having an overall out-of-plane orbital character.

In addition to taking ResPES measurements across the Cr $L_3$ ionization edge for T > $T_C$ (Fig. 4(a-b)), ARPES spectra were collected at T < $T_C$ along the ΓK axis using photon energies that have likewise been tuned across the Cr resonance (Fig. 5(a-c)). In doing so, the α band, belonging to the host $NbS_2$ compound, and two $β_{1,2}$ bands, arising from Cr intercalation, are well separated, allowing for the elemental sensitivity of ResPES to benefit from an added momentum resolution. Here, a comparison of the integrated energy distribution curves (EDCs) shown in Fig.

4(b) and Fig. 5(d) reveal a resonant enhancement of Cr 3d states occurring near $E_F$ that is independent of temperature. However, MDCs extracted from Fig. 5(a-c) at $E_F$ as a function of photon energy show an initial resonance of the $\beta_{1,2}$ bands (k ~ 0.5 Å$^{-1}$) occurring at the onset of absorption (Fig. 5(e)). This finding not only provides evidence for a Cr-derived elemental character of these bands, but also reveals a lack of hybridization with the Nb-derived $\alpha$ band that is consistent with its antibonding orbital character[32]. Given the large separation distance between Cr neighbors, the formation of dispersive $\beta_{1,2}$ bands following Cr intercalation can only occur through hybridization with Nb. This is made clear from our ResPES results, but is also indicated by the higher degree of $k_z$ dispersion exhibited by the $\beta_{1,2}$ bands as compared to $\alpha$[32]. Thus, by measuring ARPES across the Cr absorption edge, a band resolved picture of Cr-derived d states in the VB is obtained.

*Change in Spectral Weight in Proximity to $T_C$*—By revealing the presence of Cr-derived *d*-states at $E_F$, our data show that a clear separation of magnetic and itinerant DOF does not occur in Cr$_{1/3}$NbS$_2$, as the same states forming LSMs are also participating in the formation of the FS. The consequences of this finding are illustrated in the temperature dependence of the EDCs integrated over a broad momentum range ($k_{\Gamma M}$ = 0.24 Å$^{-1}$ to 0.8 Å$^{-1}$) encompassing both crossing points of the $\beta_{1,2}$ bands (Fig. 6(a)). Here, as T > 50K, the spectral weight in proximity to $E_F$ begins to drop and is transferred towards higher binding energies until T > 120K, whereupon there is no longer any change. Similarly, the temperature dependence of the $\alpha$ band appears to be less pronounced as compared to that of the $\beta$ band, but still reveals a monotonic drop of spectral weight at $E_F$, as well as a slight reduction in the separation between peaks at 100 meV and 400 meV as T > $T_C$ (Fig. 6(b)). In both cases, the spectral weight at $E_F$ changes in close proximity to $T_C$ (T = 130 K to T = 90 K), allowing for thermal broadening effects to be excluded, suggesting instead a microscopic mechanism that links electron itinerancy to the onset of ferromagnetism.

**Discussion:**

For ferromagnetism being mediated by an indirect RKKY exchange interaction, as is the case for rare earth metals like Gd, the electronic states forming the FS are distinct from those hosting LSMs[35], where the interaction between local moments and itinerant electrons is treated as a weak perturbation resulting from a FS instability[26]. Under an RKKY exchange interaction, the VB states are exchange-split by the local moment. Temperature controls the ordering of the LSMs, and affects the exchange splitting of the VB according to a behavior that is intermediate between two extreme limits. In one instance, there is no change in exchange splitting with temperature. This occurs when VB states have wavefunctions that are incoherent over the magnetic correlation length of the local moment. Such behavior is exhibited by band states that are either localized, or in systems consisting of a few magnetic impurities embedded in metals. In the opposite limit, the exchange splitting of the valence states continuously decreases with increasing temperature until vanishing all together when magnetic order is lost. This Stoner-like behavior occurs when band states are delocalized so as to remain coherent over a distance that is large compared to the magnetic correlation length of the local moment[35]. In a prototypical system described by an RKKY exchange, like Gd, VB states with different degrees of localization/delocalization exhibit behavior that spans these two extremes[36].

As this pertains to Cr$_{1/3}$NbS$_2$, despite the merging of $\beta_{1,2}$ bands as temperature is raised above $T_C$, the transfer of spectral weight away from $E_F$ as T > $T_C$ runs counter to our initial first principle, DFT calculations, showing the Stoner criterion to be well satisfied[29]. By revisiting these calculations using the same experimental structure as used by N.J. Ghimire, et al[29] with internal coordinates optimized, we focus on an underappreciated aspect of $Cr_{1/3}NbS_2$, which is how a material having such a large in-plane, Cr-Cr nearest neighbor distances (5.75 Å) can exhibit an ordering temperature as high as 131K. While evident that such a large Cr-Cr distance discounts the possibility of a direct exchange interaction, more conventional indirect exchange interactions are expected to be weak and will therefore play a role only at considerably lower energy scales. Our calculations find an energy difference between the ferromagnetic and nearest-neighbor excited antiferromagnetic state to be in excess of 30 meV/Cr, implying a $T_C \approx 116$ K under a mean-field approximation alone. Such a large energy scale for the exchange energetics is expected to dominate over any higher-order corrections, like the magnetic anisotropy[30], that arise from spin-orbit coupling in this material. While an RKKY interaction has been experimentally established as the fundamental exchange interaction in 3d intercalated TMDs [23,25], including $Cr_{1/3}NbS_2$[37], such a significant exchange energy implies the interaction between LSMs and hybridized Nb- and Cr-derived itinerant states may go beyond this perturbative interaction, originating instead from a microscopic interaction between itinerant electrons and LSMs.

The data presented here show a clear separation of magnetic and itinerant DOF does not occur, leading to the motion of itinerant electrons influencing LSMs reciprocally, and in a self-consistent fashion, as is found for other itinerant, 3d magnetic systems[38,39]. In such a case, a strong Hund's coupling establishes a connection between the onset of ferromagnetism and metallicity, as it enforces electron itinerancy on Cr sites when the spin of the majority carrier is parallel to the LSM. In this way, an on-site Hund's correlation shapes the interaction between itinerant electrons and LSM's in such a way that a synergistic and regenerative character is acquired. That is, FM order fosters electron itinerancy by a reduction of electron scattering from thermal fluctuations of the LSMs, while electron itinerancy promotes a decrease of the electron kinetic energy and FM order, overall resulting in a minimization of energy for the entire system. Hence, electron itinerancy acquires a fundamental role in the attainment of FM order, and the FM transition in $Cr_{1/3}NbS_2$ is viewed as a cooperative phenomenon driven by the mutual interaction of itinerant electrons and LSMs enforced by Hund's coupling. The concomitant increase of electron itinerancy and alignment of the LSMs provide a route for the minimization of the total energy of the system, offering a microscopic means by which important aspects of the in-plane transport can be rationalized. This includes a natural explanation for the pronounced drop of the in-plane resistivity occurring at $T_C$, along with the negative in-plane MR[28,29]. In this way, the magneto-transport properties of $Cr_{1/3}NbS_2$ are in some respect reminiscent of the colossal magnetoresistive (CMR) manganites. Both systems exhibit the same correlation between the ordering of LSMs and the onset of metallicity, and both exhibit a negative magnetoresistance, albeit quite different in order of magnitude.

In $Cr_{1/3}NbS_2$, elucidating the mechanisms that underlies both inter- and intra-layer magneto-transport acquires particular importance in light of the possible technological relevance that this material has for future spintronic applications. Our results suggest that arguments based entirely on magnetic scattering due to spin ordering in the CSL state may not fully capture the magneto-transport properties observed in this material. Seeing as our high-resolution photoemission results are sensitive to only the first few atomic layers, our findings are directly relevant to a microscopic description of the in-plane, magneto-transport properties. Nevertheless, it is unlikely that the microscopic interaction between itinerant electrons and LSMs evidenced by our data would be entirely irrelevant for the inter-layer magneto-transport properties as well. In fact, any mechanism proposed to explain the inter-layer MR must be rooted in a Hamiltonian that takes into account the interaction of itinerant electrons with the LSMs. Thus, despite the long length-scales of the CSL, our results highlight the relevance of the electronic structure in a description of the magneto-transport properties in $Cr_{1/3}NbS_2$, as underscored by the strong Hund's coupling between itinerant electrons and LSMs. For this reason, we hope this work encourages future investigations into other 3d intercalated TMDs, where the interaction between LSMs and itinerant electrons extends beyond a rigid band interpretation and will be relevant for shaping anisotropic functionalities in this class of materials.

In closing, electronic structure measurements of $Cr_{1/3}NbS_2$ taken above and below the helimagnetic transition temperature, reveal a clear separation of magnetic and itinerant DOF does not occur, as the same states forming LSMs are also participating in the formation of the FS. An anomalous increase of spectral weight in proximity to the Fermi level ($E_F$) as temperature is lowered below $T_C$ marks behavior that is inconsistent with conventional itinerant ferromagnets, and is rationalized on the basis of first principle density functional theory (DFT) calculations to result from a strong exchange (Hund's) interaction between itinerant electrons and LSMs on the Cr sites. Our results suggest that arguments based entirely on magnetic scattering due to spin ordering in the CSL state may not fully capture the magneto-transport properties observed in this material, as the electronic structure is clearly shown to play a non-trivial role as temperature is tuned across $T_C$.

## Methods

*Crystal Growth—P*olycrystalline $Cr_{1/3}NbS_2$ samples were grown by heating stoichiometric ratios of Cr, Nb and S to 950 °C for one week. Single crystal growth was carried out under chemical vapor transport using 0.5 g Iodine transport agent per 3 g $Cr_{1/3}NbS_2$. 5 mm x 5 mm plate-like crystals oriented along (001) formed across a 100 °C (950 °C – 850 °C) temperature gradient of the transport tube. Due to the deleterious effect that Cr disorder has on the appearance of helimagnetic ordering in this material[40], x-ray and low energy electron diffraction measurements were used to confirm a $P6_322$ space group showing ($\sqrt{3}$ x $\sqrt{3}$) Cr ordering[29,32]. Sample quality was further verified with superconducting quantum interference device (SQUID) magnetometry (Fig. 1(b)), which reveals a prominent kink at $T_C$ = 131K, indicative of helimagnetic ordering in this sample batch.

*Photoemission Spectroscopy—T*emperature dependent ARPES experiments were carried out on $Cr_{1/3}NbS_2$ single crystals cleaved in-situ on Beamline 10.0.1 of the Advanced Light Source (ALS) and the Advanced Photoelectric Effect beamline at the Elettra Synchrotron Facility. The total instrumental energy resolution ranged from 15 to 30 meV, while an angular resolution of ±0.5°, gives a momentum resolution of < 0.06 Å$^{-1}$ for the photon energies used in these experiments (hν = 40 eV and 48 eV). ResPES and Res-ARPES experiments were performed above and below $T_C$ on the Beamline for Advanced diCHroism (BACH) at Elettra with a total instrumental energy resolution being better than 300 meV.

*Ab initio Calculations—F*irst principles DFT calculations were carried out using the linearized augmented plane-wave density functional theory code WIEN2K[41] in the generalized gradient approximation of Perdew, Burke and Ernzerhof[42]. Respective sphere radii of 2.01, 2.33 and 2.37 Bohr were used, for S, Cr and Nb, with a value for the product of the smallest sphere radius (S) and the largest plane-wave expansion vector being set to $RK_{max}$ = 8.0. Calculations of the magnetic properties, performed with and without spin-orbit coupling, were carried out under the internal coordinates of the structure being relaxed in the ferromagnetic state. A minimum of 800 k-points in the full Brillouin zone were used for the ferromagnetic calculations, while this number was proportionally reduced for the 2 x 2 x 1 supercell of the planar antiferromagnetic state.

A comparison of DFT energetics with and without spin-orbit coupling is made between a ferromagnetic and two possible anti-ferromagnetic states denoted as $AFM_1$ and $AFM_2$. These states represent either anti-aligned Cr-Cr pairs along the c-axis ($AFM_1$), or four of six anti-aligned Cr-Cr neighboring pairs in the ab-plane ($AFM_2$). From our calculation, we find an energy difference in meV/Cr for $AFM_1$ and $AFM_2$ relative to the ferromagnetic state that shows DFT can describe the normal (i.e. non-Dzyaloshinskii-Moriya) exchange interaction without the need to consider spin-orbit coupling (Table 1). This < 1 meV/Cr difference is fully consistent with the high (131 K) magnetic ordering temperature found in this material.

|  | $AFM_1$/FM | $AFM_2$/FM |
| --- | --- | --- |
| Spin-Orbit (meV/Cr) | 34.9 | 30.8 |
| No Spin-Orbit (meV/Cr) | 35.0 | 29.9 |
| Difference (meV/Cr) | -0.1 | +0.8 |

Table 1: *Comparison of First Principle Calculations with and without Spin-Orbit Coupling* – Calculated energy in meV/Cr of two antiferromagnetic (AFM) excited states relative to the ferromagnetic (FM) ground states with and without spin-orbit coupling.

**Acknowledgements**
This work was performed, in part, at the Advanced Light Source, a DOE Office of Science User Facility under contract no. DE-AC02-05CH11231. NS acknowledges support from the Center for Integrated Nanotechnologies at Los Alamos National Laboratory (LANL), a U.S. Department of Energy, Office of Basic Energy Sciences user facility.

**Data Availability**
All data presented in this manuscript can be made available upon reasonable request to corresponding authors, NS and NM.

**Author Contributions**
Samples were provided by L. L. D. S. and D. G. M. Temperature dependent ARPES experiments were performed by N. S. and N. M. with the help of S. -K. M., A. V. F., P. K. D., I. V., and J. F. ResPES and ResARPES experiments were carried out by N.S. with the help of F. B., I. P., and S. N. N.S. and P. V. analyzed the data, with input from N.M. and D.S.P., who provided a*b initio* calculations. N.S. P.V. and N.M. wrote the manuscript with input from all authors.

**Competing Interests**
The authors declare no competing interests

**Figures**

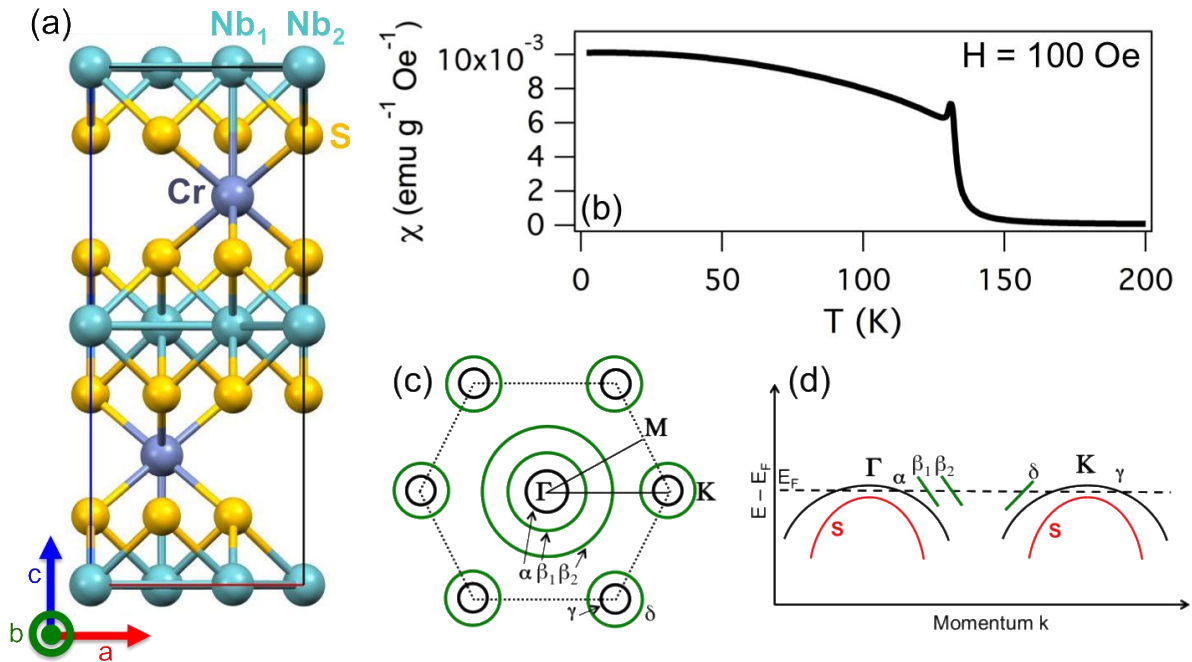

Fig. 1: *Crystal and Electronic Structure of $Cr_{1/3}NbS_2$ in the Helimagnetic State*—$Cr_{1/3}NbS_2$ (a) unit cell, (b) magnetic susceptibility, (c) Fermi surface and (d) electronic band dispersion along the ΓK direction for temperatures below the helimagnetic transition temperature, $T_C$. The unit cell, defined by lattice constants $a = b = 5.741$Å and $c = 12.101$Å, contains 20 atoms, with twelve S atoms occupying the general site, and six Nb atoms having two inequivalent positions. Helimagnetic order is verified by the presence of a prominent kink in the magnetic susceptibility at 131 K as measured within a 100 Oe external magnetic field. The $NbS_2$-derived hole pockets at Γ and K are shown in black, while additional bands arising from Cr intercalation are shown in green. The three dispersive bands, labeled $\alpha$, $\beta_1$, and $\beta_2$, found near zone center are the focus of this experimental work.

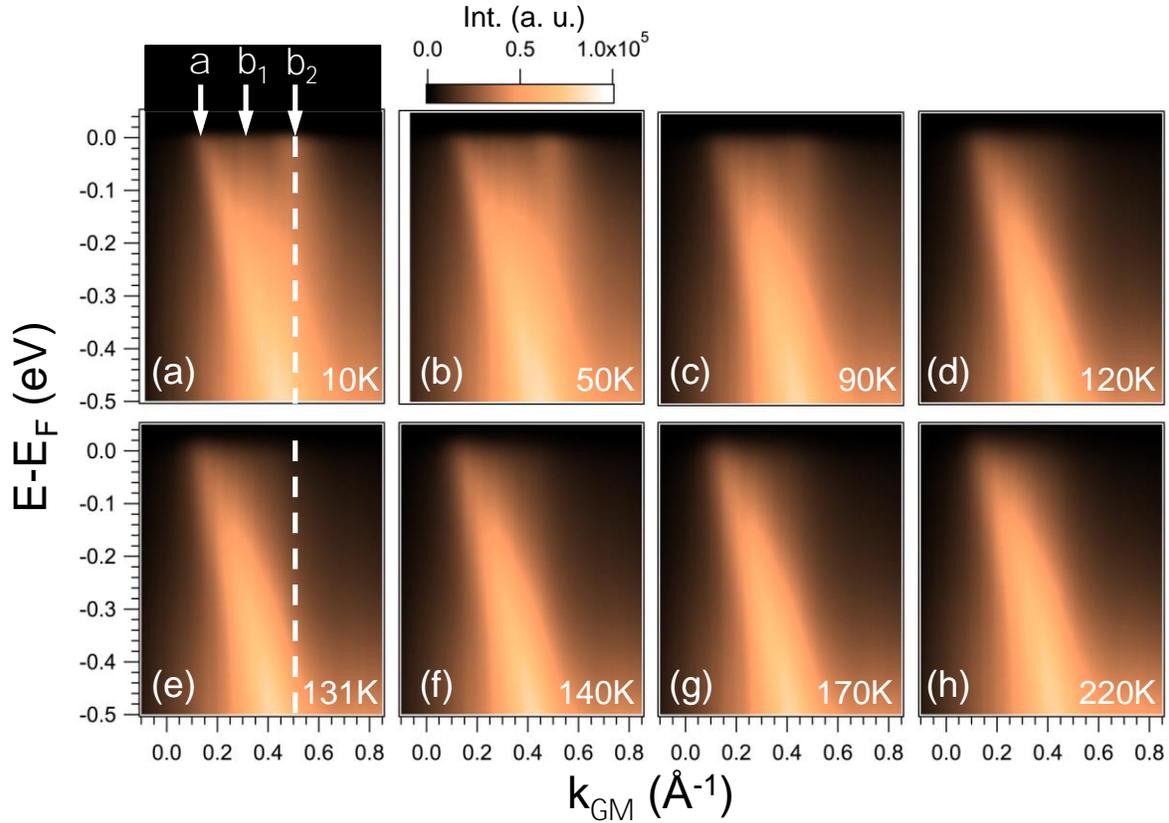

Fig. 2: *Temperature-Dependent Angle-Resolved Photoemission*—Temperature-dependent angle-resolved photoemission spectra obtained along the ΓM direction at (a) 10 K (b) 50 K (c) 90 K (d) 120 K (e) 131 K (f) 140 K (g) 170 K and (h) 220 K using π-polarized photons having an energy, hν = 48 eV. Here, band crossing points are denoted by white arrows in (a), while the dashed line in (a) and (e) denotes the presence and absence of a $\beta_2$ band crossing as temperature is raised above the helimagnetic transition temperature.

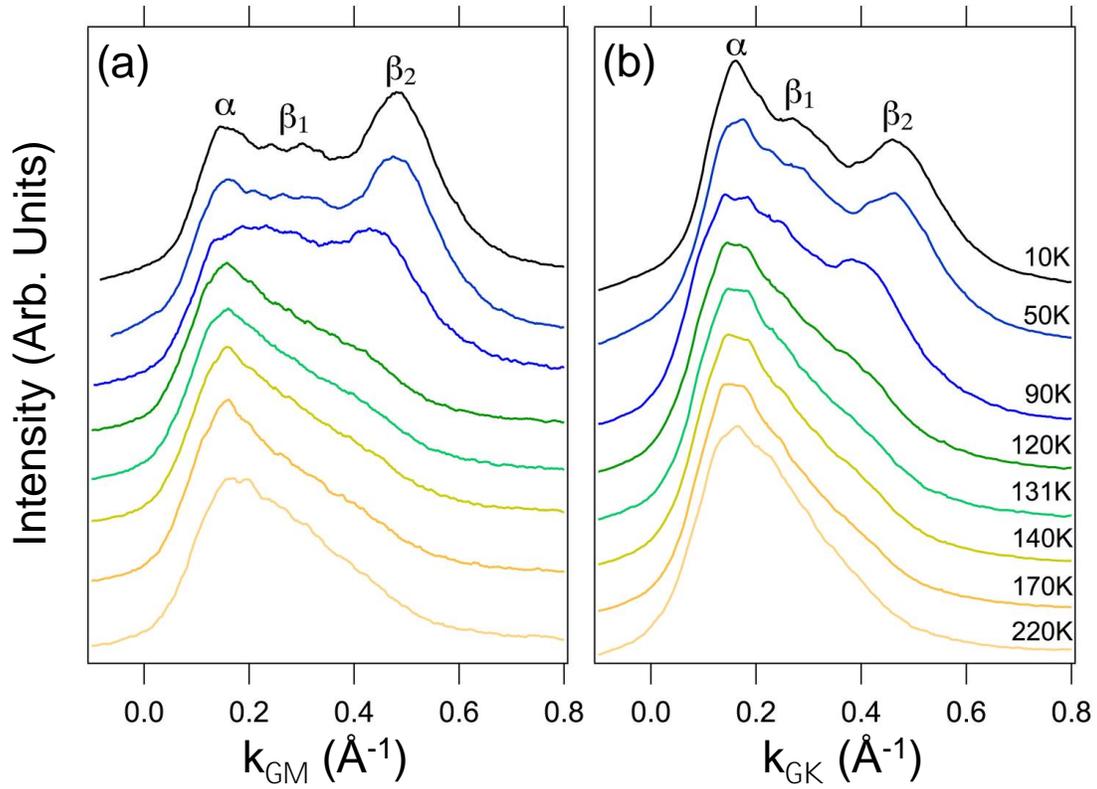

Fig. 3: *Temperature Dependence of Band Crossing Points*—Momentum distribution curves extracted at the Fermi level from spectra measured along (a) ΓM and (b) ΓK using photons of energy hν = 48 eV. Note the emergence of a split $\beta_{1,2}$ band for temperatures below 120 K, and a loss of spectral weight along both high symmetry axes for temperatures below the helimagnetic transition temperature.

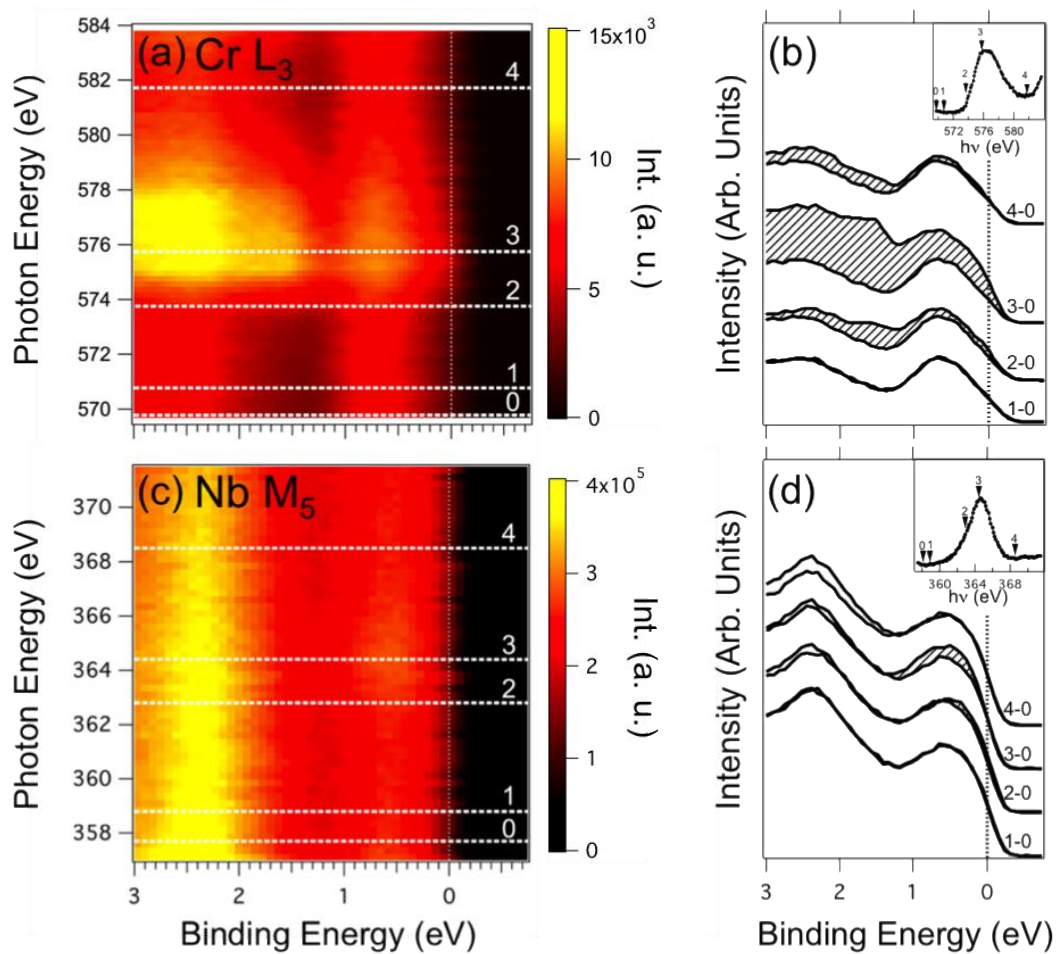

Fig. 4: *Identifying Nb- and Cr- Derived States in the Valence Band*—Resonant photoemission spectra measured across the (a-b) Cr $L_3$ and (c-d) Nb $M_5$ absorption edges using linear horizontal photon polarization. Intensity maps ((a) and (c)) and selected photoemission spectra generated across the (b) Cr and (d) Nb x-ray absorption edge (XAS) shown in the inset. Spectra denoted 1 – 4 in panel (a) and (c) are taken through the Cr and Nb resonance respectively. Hatched area in (b) and (d) denotes the difference between spectra taken on- (1-4) and off- (0) resonance.

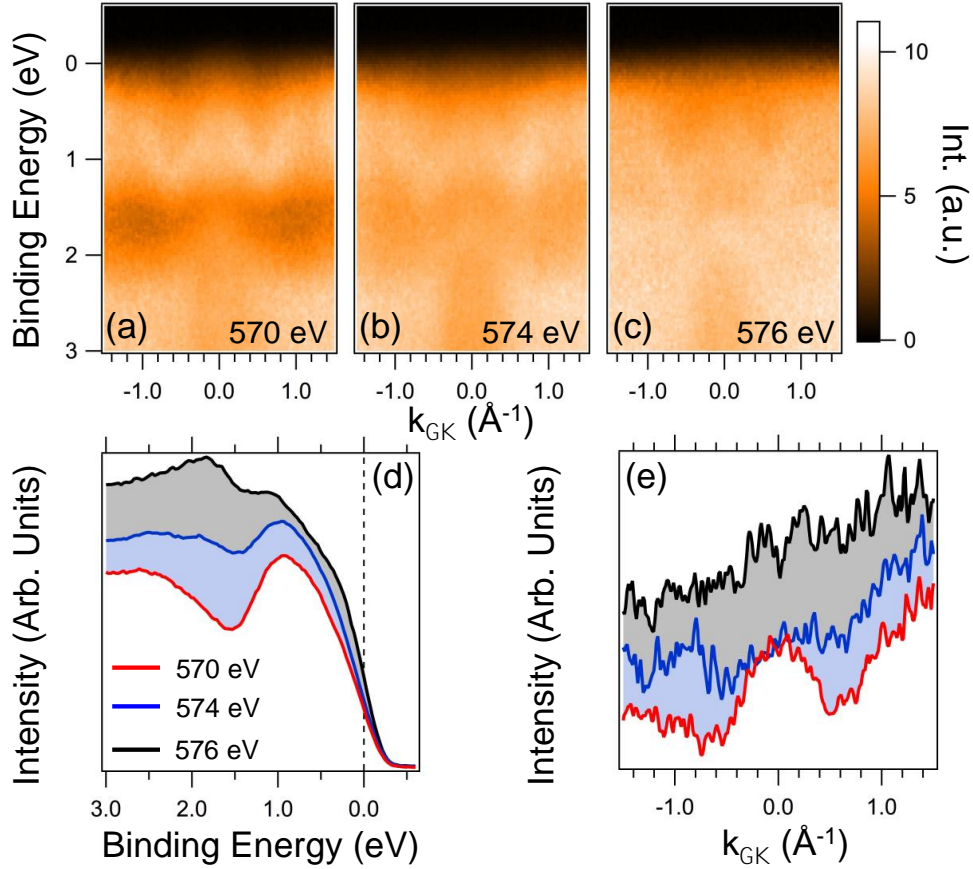

Fig 5: *Angle-Resolved Photoemission Across the Cr Resonance*—Angle-resolved photoemission spectra measured below the helimagnetic transition temperature (110K) using a photon energy tuned (a) off the Cr $L_3$ resonance (hν = 570 eV), (b) at the onset of Cr absorption (hν = 574 eV) and (c) at the maximum of Cr absorption (hν = 576 eV), where the contribution due to photon momentum is taken into account. An angle of 60° is made between the polarization vector of the incoming light and the crystal surface plane, resulting in a dominant component perpendicular to the sample surface. (d) Integrated energy dispersive curves taken over a momentum range, $\Delta k = \pm 0.5$ Å$^{-1}$ with respect to Γ, encompassing the α and two $\beta_{1,2}$, bands. (e) Resonant momentum distribution curves obtained by integrating ±50 meV about the Fermi level. For the resonance tuned on the maximum of Cr absorption, a featureless MDC is observed, indicating various scattering channels have opened across the Fermi surface.

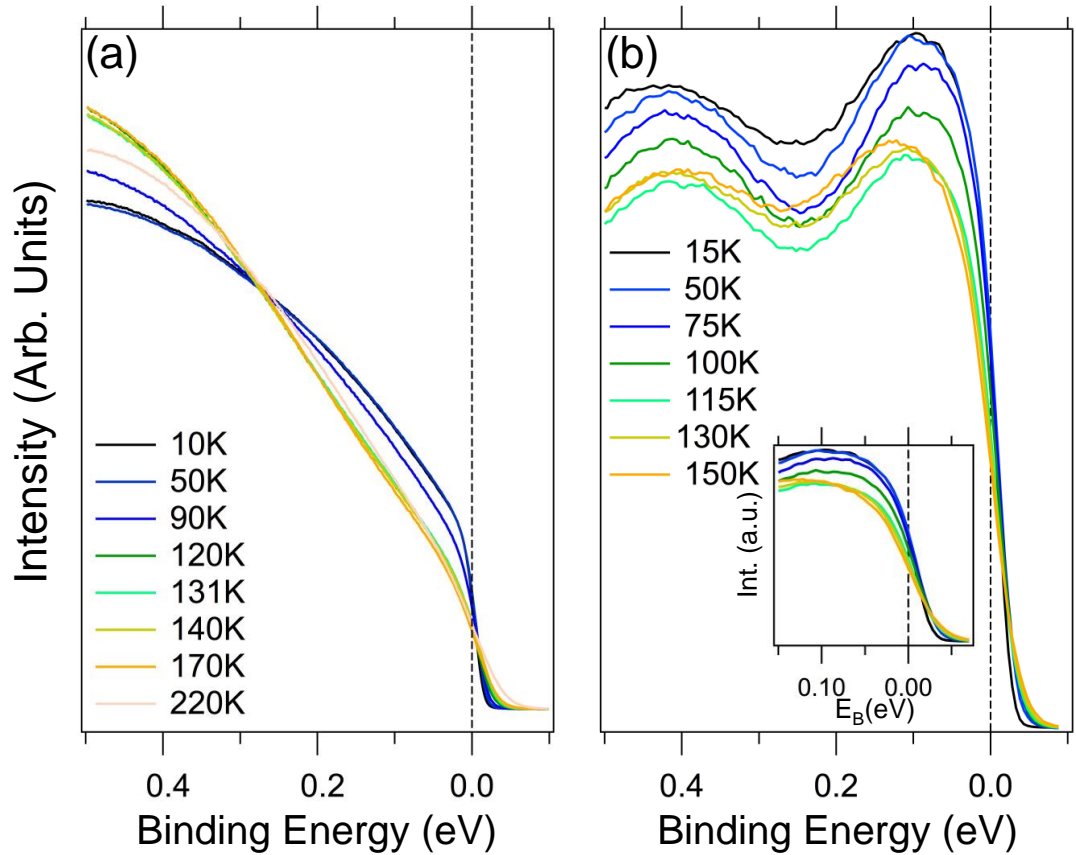

Fig. 6: *Changes in Spectral Weight near the Fermi Level Across the Helimagnetic Transition*— (a) Energy distribution curves normalized to photon flux and integrated from $k_{\Gamma M}$ = 0.24 Å$^{-1}$ - 0.8 Å$^{-1}$ (b) Energy distribution curves extracted at the Fermi crossing of the α band as a function of temperature from a separate temperature-dependent angle-resolved photoemission study used to confirm the findings in (a). Note the monotonic suppression of spectral weight as the temperature is increased above the helimagnetic transition temperature. This behavior is more clearly highlighted by the inset in (b) showing a change in spectral weight occurring within 150 meV from the Fermi level.